\documentclass[reprint,amsmath,amssymb,aps,longbibliography,showpacs]{revtex4-1}

\usepackage{graphicx}
\usepackage{dcolumn}
\usepackage{bm}
\usepackage{CJK}
\usepackage{subfigure}
\usepackage{color}
\usepackage{epstopdf}

\begin{document}
\begin{CJK*}{GBK}{song}

\preprint{APS/123-QED}
\title{Floquet control of the gain and loss in a $\mathcal{PT}$-symmetric optical coupler}

\author{Yi Wu$^{1}$}
\author{Bo Zhu$^{2}$}
\author{Shufang Hu$^{3}$}
\author{Zheng Zhou$^{4}$}
\altaffiliation{zhouzheng872@163.com}
\author{Honghua Zhong$^{3}$}
\altaffiliation{hhzhong115@163.com}

\affiliation{$^{1}$Department of Physics, Engineering University of CAPF, Xi'an 710086, China}
\affiliation{$^{2}$School of Physics and Astronomy, Sun Yat-Sen University(Zhuhai Campus), Zhuhai 519082, China}
\affiliation{$^{3}$Institute of Mathematics and Physics, Central South  University of Forestry and Technology, Changsha 410004, China}
\affiliation{$^{4}$Department of Physics, Hunan Institute of Technology, Hengyang 421002, China}

\begin{abstract}
Controlling the balanced gain and loss in a $\mathcal{PT}$-symmetric system is a rather challenging task.
Utilizing Floquet theory,
we explore the constructive role of periodic modulation in
controlling the gain and loss of a $\mathcal{PT}$-symmetric optical coupler. It is found that the gain and loss of the system can be manipulated by applying a periodic modulation. Further, such an original non-Hermitian system can even be modulated into an effective Hermitian system derived by the high-frequency Floquet method. Therefore, compared with other $\mathcal{PT}$ symmetry control schemes, our protocol
can modulate the unbroken $\mathcal{PT}$-symmetric range to a wider parameter region. Our results provide a promising approach for controlling the gain and loss of a realistic system.
\begin{center}
\begin{minipage}{14.0cm}
\begin{minipage}[t]{1.8cm}{\bf Keywords}\end{minipage}
$\mathcal{PT}$ symmetry, periodic modulation, optical coupler
\end{minipage}\par\vglue8pt
\end{center}
\end{abstract}

\pacs{11.30.Er, 42.81.Qb, 42.82.Et}
\maketitle
\section{Introduction}

Parity-time ($\mathcal{PT}$) symmetry, which is the invariance under simultaneous parity and time reversal transformation, plays an important role in non-Hermitian (NH) quantum mechanics and optics \cite{mahaux,konotop,moiseyev2011}, with potential applications in optical beam engineering, mode conversion, image processing, laser mode selection, asymmetric transmission, and other fields \cite{hodaei2014,
feng2014,feng2013,peng2014,naza2014,xiong2015,
longhi2014,karta2014,xiong2016,hoda2015,ywu,hjing,
karta2016,wu2016,wang16}.
The $\mathcal{PT}$ symmetry of a Hamiltonian is defined as $\mathcal{PT}H=H \mathcal{PT}$ with the parity operator ($\hat{P}: \hat{x}\rightarrow -\hat{x}, \hat{p}\rightarrow -\hat{p}$) and time-reversal operator ($\hat{T}: \hat{x}\rightarrow \hat{x}, \hat{p}\rightarrow -\hat{p}, i\rightarrow -i, t\rightarrow -t$), where $\hat{x}$ is the position operator, $\hat{p}$ is the momentum operator, and $t$ denotes time.
As the operators $\mathcal{PT}$ and $H$ may share common eigenfunctions, a broad class of non-Hermitian $\mathcal{PT}$-symmetric Hamiltonians can still have entirely real eigenvalue spectra.
However, depending on the values of the gain and loss parameters, the $\mathcal{PT}$-symmetry may be spontaneously broken \cite{bender,bender2}, and then the eigenvalues become complex.
In the last few years, $\mathcal{PT}$-symmetry and $\mathcal{PT}$-symmetry-breaking have been observed in several optical experiments \cite{ruter,guo,bersch,dietz}.

An important
issue in a $\mathcal{PT}$-symmetric system is the ability to control and
tune the $\mathcal{PT}$ phase transition. One possibility is obviously
to vary the level of gain and loss in the system; however,
this is a rather challenging task, as the gain and loss should
remain balanced and thus must be tuned simultaneously.
In recent years, it has been proposed that periodic modulations can be used to control $\mathcal{PT}$-symmetry in two-state systems \cite{Moiseyev,lee,lian,jog,gong,zhouz}.
It has been found that pseudo-$\mathcal{PT}$-symmetry may appear whether or not the original system is $\mathcal{PT}$-symmetric, and spontaneous $\mathcal{PT}$-symmetry-breaking sensitively depends on the modulation parameters \cite{lee}. These methods are based on the high-frequency Floquet method to rescale the coupling strength. Therefore, the unbroken $\mathcal{PT}$-symmetric range only can be modulated narrow.
Another means of control is to geometrically twist the fiber \cite{longhi16}, which can introduce additional Peierls phases in the coupling constants among
the fiber, and thus the transition from unbroken to broken $\mathcal{PT}$-symmetric phases can be conveniently controlled.
Although there are several studies on the manipulation of $\mathcal{PT}$-symmetry, how to control the balanced gain and loss is still lacking.

In this paper, we investigate how to control the balanced gain and loss in a $\mathcal{PT}$-symmetric optical coupler by periodically modulating the coupling strength between two waveguides. Using the
high-frequency Floquet method, the modulated system is effectively described by an effective averaged system whose gain and loss can be
modulated by adjusting the modulation amplitude or frequency. Such an original non-Hermitian system can even be modulated into an effective Hermitian system.
The spontaneous $\mathcal{PT}$-symmetry-breaking transition is analytically derived and is well consistent with the numerical simulation. It is revealed that the unbroken $\mathcal{PT}$-symmetric range can be modulated to a wider parameter region.

The structure of this article is as follows.
In Section II, we explore the constructive role of periodic modulation in controlling the gain and loss of a $\mathcal{PT}$-symmetric optical coupler.
In Section III, we study the Floquet modulation of $\mathcal{PT}$-symmetry and the corresponding dynamics.
In Section IV, the possibility of experimentally observing
our theoretical predictions is discussed.
In the last section, we briefly summarize our results.

\section{Control of the gain and loss}
We consider a periodically modulated linear $\mathcal{PT}$-symmetric coupler that is described by the following coupled-mode equation:
\begin{eqnarray} \label{twomodeequation}
i\frac{dc_1}{dz}&=&
\Big[\frac{\upsilon}{2}+\frac{F\cos(\omega z)}{2}\Big]c_2+i \gamma c_1,
\nonumber \\
i\frac{dc_2}{dz}&=&
\Big[\frac{\upsilon}{2}+\frac{F\cos(\omega z)}{2}\Big]c_1-i \gamma c_2.
\end{eqnarray}
Here, $c_1(z)$ and $c_2(z)$ are the complex amplitudes and $z$ is the coordinate in the propagation direction. Parameter $\upsilon$ is the interchannel coupling strength, $\gamma$ is the gain or loss
strength, $F$ is the modulation amplitude, and $\omega$ is the modulation frequency. Obviously, by defining the parity operator as $\hat{P}$, which interchanges the two channels labeled by 1 and 2, and the time-reversal operator as $\hat{T}$: $i\rightarrow-i$, $z\rightarrow-z$, which reverses the propagation direction, the Hamiltonian $\hat{H}$ of system (\ref{twomodeequation}) is $\mathcal{PT}$ symmetric because $\hat{\mathcal{P}}\hat{\mathcal{T}}\hat{H}
\hat{\mathcal{P}}\hat{\mathcal{T}}=\hat{H}$. To simplify, below
we only consider the case of $\upsilon>0$, as the system is invariant under the transformation $c_{2}\rightarrow-c_{2},\upsilon\rightarrow-\upsilon$ and all the parameters are dimensionless throughout this paper.

\begin{figure}[htp]
\center
\includegraphics[width=3.4in]{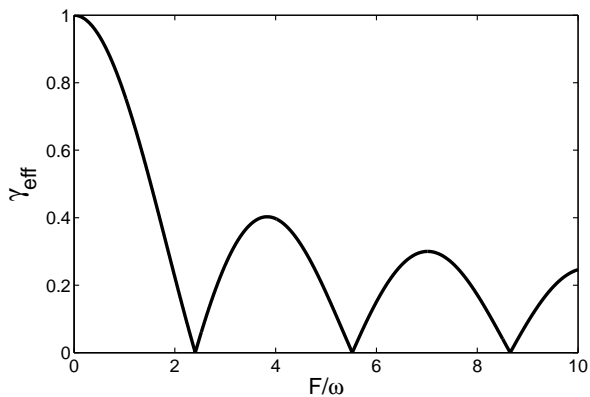}
\caption{Parametric dependence of the effective gain and loss $|\gamma_{eff}|$ for $\gamma=1$ and $\omega=10$.}  \label{fig1}
\end{figure}

By making use of the linear combinations $\psi_{1}=c_{1}-c_{2}$ and $\psi_{2}=c_{1}+c_{2}$, we can rewrite Eq.(\ref{twomodeequation}) as
\begin{eqnarray} \label{twomodeequation2}
i\frac{d}{dz}\left( \begin{array}{c}
\psi_1 \\
\psi_2 \\
\end{array}
\right)=\left(\begin{array}{cc}
-\frac{\upsilon+F\cos(\omega z)}{2} & i \gamma
\\  i\gamma & \frac{\upsilon+F\cos(\omega z)}{2}
\end{array}\right)\left( \begin{array}{c}
\psi_1 \\
\psi_2 \\
\end{array}
\right).
\end{eqnarray}
Under the condition $\gamma \ll \max[\omega,\sqrt{|F|\omega}]$,
one can perform high-frequency Floquet analysis.
Then, by introducing the transformation
\begin{eqnarray}
\psi_{1}&=&\psi_{1}^{'}\exp[i\frac{F}{2\omega}\sin(\omega z)],\nonumber
\\ \psi_{2}&=&\psi_{2}^{'}\exp[-i\frac{F}{2\omega}\sin(\omega z)]
\end{eqnarray}
and averaging the high-frequency terms, one can obtain the effectively unmodulated system
\begin{eqnarray} \label{twomodeequation3}
i\frac{d}{dz}\left( \begin{array}{c}
\psi_1^{'} \\
\psi_2^{'} \\
\end{array}
\right)=\left(\begin{array}{cc}
-\frac{\upsilon}{2} & i\gamma_{eff}
\\  i\gamma_{eff}^{*} & \frac{\upsilon}{2}
\end{array}\right)\left( \begin{array}{c}
\psi_1^{'} \\
\psi_2^{'} \\
\end{array}
\right)
\end{eqnarray}
with the rescaled gain and loss parameter
\begin{eqnarray}
\gamma_{eff}=\gamma \Sigma^{\infty}_{k=-\infty}(ie^{i\omega t})^{k}J_{k}(F/\omega).
\end{eqnarray}
Here, $J_{k}(F/\omega)$ are ordinary Bessel functions. Therefore, the modulus of $\gamma_{eff}$ depends on the values of $F/\omega$, and can
change from $\gamma$ to zero at some specific values of $F/\omega$ (such as $F/\omega\simeq$ 2.4 and 5.52), as shown in Fig. \ref{fig1}. When $|\gamma_{eff}|=0$, the effective system corresponds to a Hermitian system. Surprisingly, in contrast to our conventional understanding, we find that the gain and loss of system can be manipulated by applying a periodic modulation, and even an original non-Hermitian system can be modulated into an effective Hermitian system. As is well known, the change of the gain and loss parameter is connected to the transition from unbroken to broken $\mathcal{PT}$ symmetry. Our results imply that the unbroken $\mathcal{PT}$-symmetric range can be modulated wide, which is different from previous results \cite{lee,lian,jog,gong}, where the unbroken $\mathcal{PT}$-symmetric range only can be modulated  narrow.

\section{Manipulation of $\mathcal{PT}$ symmetry and dynamics}
By diagonalizing the Hamiltonian for effective model (\ref{twomodeequation3}), two eigenvalues can be easily determined by
\begin{eqnarray} \label{eigenequation}
\varepsilon=\pm|\gamma_{eff}|\sqrt{(\frac{\upsilon}{2|\gamma_{eff}|})^2-1)}.
\end{eqnarray}
Obviously, dependent on the values of $\upsilon/2|\gamma_{eff}|$, the
two eigenvalues can be real or complex. The two eigenvalues are real
if $\upsilon>2|\gamma_{eff}|$, and they become complex if $\upsilon<2|\gamma_{eff}|$. Therefore, $\upsilon=2|\gamma_{eff}|$ is the critical point for the phase
transition between the real and complex spectra in the effective
system, which corresponds to original system (\ref{twomodeequation2})
under high-frequency modulation. The spontaneous
$\mathcal{PT}$-symmetry-breaking transition takes place in effective
model (\ref{twomodeequation3}) when the imaginary part $|\text{Im}(\varepsilon)|$ changes
from zero to nonzero.
Note that because $|\gamma_{eff}|$ can be modulated from $\gamma$ to 0, the unbroken $\mathcal{PT}$-symmetric range can be modulated wide by tuning the amplitude of periodic modulation for a fixed $\upsilon$.

\begin{figure}[htp]
\center
\includegraphics[width=3.4in]{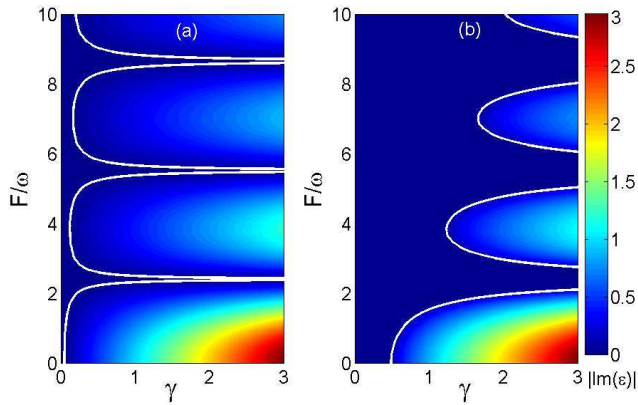}
\caption{(Color online) The imaginary parts of the quasienergies $|Im(\varepsilon)|$ as a function of $F/\omega$ and $\gamma$ for different coupling strengths $\upsilon=0.1$ (a) and $\upsilon=1$ (b). The other parameter is fixed as $\omega=10$. The white curves are the boundary ($\gamma_{eff}=\upsilon/2$) between $|Im(\varepsilon)|=0$ and $|Im(\varepsilon)|\neq0$.} \label{fig2}
\end{figure}

To show the parametric dependence of $|Im\varepsilon|$, in Figs. \ref{fig2}(a) and \ref{fig2}(b), we show $|Im(\varepsilon)|$ as a function of $F/\omega$ and $\gamma$ for both a small coupling strength $\upsilon=0.1$ and large coupling strength $\upsilon=1$ with $\omega=10$. First, it is clear
that, compared with a $\mathcal{PT}$-symmetric system with no modulation, the region of $\mathcal{PT}$ symmetry in our modulated system can obviously be manipulated to be wider by tuning $F/\omega$, which is different to previous results \cite{lee,jog,gong}. Second, near a minimum of $|\gamma_{eff}|$ such as $F/\omega\simeq 2.4$, despite the size of $\upsilon$ and $\gamma$, there always exists a completely real quasienergy spectrum.

\begin{figure}[htp]
\center
\includegraphics[width=3.6in]{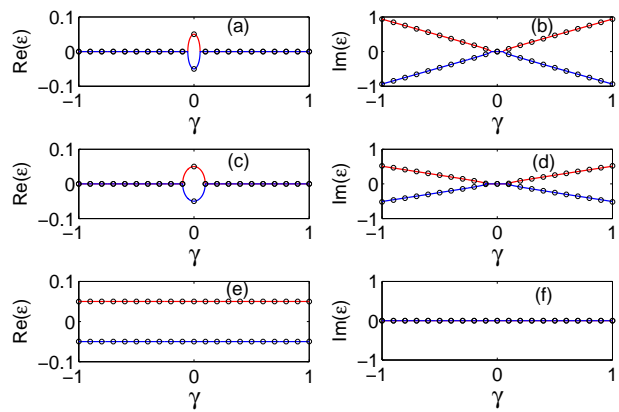}
\caption{(Color online) The real parts $Re(\varepsilon)$ and the imaginary parts $Im(\varepsilon)$ of the complex quasienergies as a function of $\gamma$ for different driving amplitudes $F=5$ (top row), $F=15$ (middle row) and $F=24$ (bottom row). The solid lines correspond to analytical results given by the formula (\ref{eigenequation}) for the effective model (\ref{twomodeequation3}) and the circles correspond to numerical results obtained from the original model (\ref{twomodeequation}). The other parameters are $\upsilon=0.1$ and $\omega=10$.} \label{fig3}
\end{figure}

According to the Floquet theorem, one can use a numerical method
to calculate the Floquet states and their quasienergies for arbitrary
modulation amplitude and frequency. The Floquet states of
modulated system (\ref{twomodeequation}) satisfy $c_j(z)=e^{-i\varepsilon z}\tilde{c}_j(z)$ for $j=1,2$, where propagation constant $\varepsilon$ is designated as
the quasienergy, and complex amplitudes $\tilde{c}_j(z)$ are periodic with modulation period
$T=2\pi/\omega$.
The quasienergies and amplitudes of
modulated system (\ref{twomodeequation}) are
then given by
\begin{eqnarray} \label{Fequation}
\mathcal{F}\tilde{c}_j(z)=\varepsilon \tilde{c}_j(z), \ \ \ j=1,2,
\end{eqnarray}
with the Floquet operator
\begin{eqnarray} \label{fq}
\mathcal{F}=-i\frac{d}{dz}+H(z).
\end{eqnarray}
To verify the validity of the above high-frequency Floquet analysis,
we compare the numerical quasienergies obtained from original model
(\ref{twomodeequation}) and analytical formula (\ref{eigenequation})
obtained from effective model (\ref{twomodeequation3}). As an example, we show the real parts $Re(\varepsilon)$ and imaginary parts $Im(\varepsilon)$ of quasienergies $\varepsilon$ as a function of $\gamma$ for different driving amplitudes $F=5$, $15$, and $24$
in Fig. \ref{fig3}. The other parameters are chosen as $\upsilon=0.1$ and $\omega=10$. Obviously, in the high-frequency
regime $\gamma \ll \max[\omega,\sqrt{|F|\omega}]$ the analytical (solid lines)
and numerical (circles) values for the quasienergies $\varepsilon$ are in good
agreement. This clearly shows that below the critical point ($|\gamma_{eff}|<\upsilon/2$), both the numerical and analytical results confirm the entirely real quasienergy spectrum, and the transition from a completely real quasienergy
spectrum ($|Im(\varepsilon)|=0$) to a complex spectrum  ($|Im(\varepsilon)|\neq0$) can take place when $\gamma$ increases.            In particular, in contrast to a $\mathcal{PT}$-symmetric system without periodic modulation, the spontaneous $\mathcal{PT}$-symmetry-breaking transition can be manipulated by tuning the modulation parameter. For small $F$, for example $F=5$ in Fig. \ref{fig3}(b), one clearly sees that there exists a narrow real quasienergy spectrum region, and the spontaneous $\mathcal{PT}$-symmetry-breaking transition takes place at $\gamma\doteq0.05$.
For a bigger $F$,
such as $F=15$, the region of the real quasienergy spectrum broadens and the spontaneous $\mathcal{PT}$-symmetry-breaking transition occurs at $\gamma_c\doteq0.1$ (see
Fig. \ref{fig3}(d)). The quasienergy spectrum is always completely real regardless of the size of $\gamma$ for $F=24$, which is equivalent to a Hermitian case, as shown in Fig. \ref{fig3}(f).

\begin{figure}[htp]
\center
\includegraphics[width=3.6in]{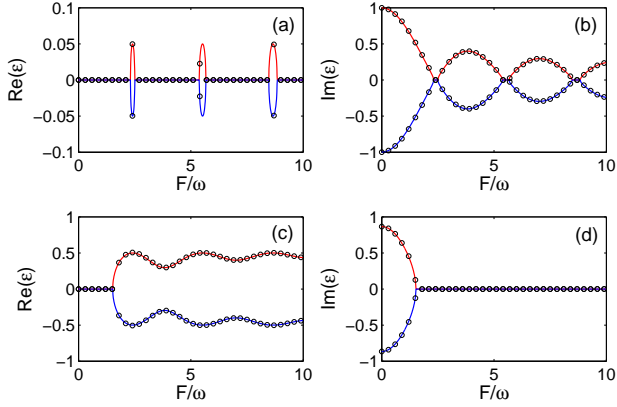}
\caption{(Color online) The real parts $Re(\varepsilon)$ and the imaginary parts $Im(\varepsilon)$ of the complex quasienergies as a function of $F/\omega$ for two different coupling strengths $\upsilon=0.1$ (top row) and $\upsilon=1$ (bottom row). The solid lines correspond to analytical results given by the formula (\ref{eigenequation}) for the effective model (\ref{twomodeequation3}) and the circles correspond to numerical results obtained from the original model (\ref{twomodeequation}). The other parameters are chosen as $\gamma=1$ and $\omega=10$.} \label{fig4}
\end{figure}

To understand this concept from another angle, we also show
the quasienergy dependence on modulation parameter $F/\omega$
for different parameters $\upsilon$ in Fig. \ref{fig4}. The other parameters are chosen as $\gamma=1$ and $\omega=10$.
As shown in Figures \ref{fig4}(a) and \ref{fig4}(b), for a small coupling strength $\upsilon=0.1$, the $\mathcal{PT}$ symmetry can spontaneously break when parameter $F/\omega$ changes. If the coupling strength is increased to $\upsilon=1$, a completely real quasienergy spectrum always appears when modulation parameter $F/\omega$ exceeds a critical value. These results show that the region of $\mathcal{PT}$ symmetry can be manipulated wider by tuning $F/\omega$ in the case of bigger coupling strength for a fixed parameter $\gamma$.

\begin{figure}[htp]
\center
\includegraphics[width=3.6in]{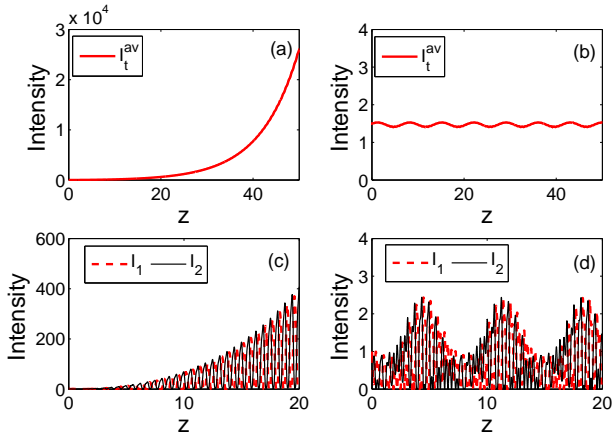}
\caption{(Color online) Intensity evolution from Eq. (\ref{twomodeequation}) for the initial
state of $c_1(0)=1$ and $c_2(0)=0$. Upper row: long-distance time averaged total intensity evolution for (a) $F/\omega=1.52$ and (b) $F/\omega=2.0$.
Lower row: short-distance intensity evolution for (c) $F/\omega=1.52$
and (d) $F/\omega=2.0$. The other parameters are chosen as $\upsilon=1$,  $\gamma=1$ and $\omega=10$.} \label{fig5}
\end{figure}

In the following, through numerical integration, we analyze the light propagation in
unbroken and broken $\mathcal{PT}$-symmetric parameter regions.
To do this, we define the two intensities in
coupled-mode system (\ref{twomodeequation}) as $I_{1}(z)=|c_{1}(z)|^{2}$ and $I_{2}(z)=|c_{2}(z)|^{2}$, the total
intensity as $I_{t}(z)=I_{1}(z)+I_{2}(z)$, and the time-averaged
total intensity as $I_t^{av}(z)=\frac{1}{T_s}\int_z^{z+T_s} I_{t}(\tilde{z}) d{\tilde{z}}$ with $T_s=2\pi/|Re(\varepsilon_2)-Re(\varepsilon_1)|$ being the two real parts of $\varepsilon$.
In Fig. \ref{fig5}, for $\upsilon=1$, $\gamma=1$, and $\omega=10$, we show the evolution in intensity from initial
states $c_1(0)=1$ and $c_2(0)=0$ for $F/\omega=1.52$ (broken $\mathcal{PT}$-symmetric parameter, see Fig. \ref{fig2}(b)) and $F/\omega=2.0$ (unbroken $\mathcal{PT}$-symmetric parameter, see Fig. \ref{fig2}(b)). We can see that
the light propagation sensitively
depends upon the $\mathcal{PT}$ symmetry. Stationary light
propagations of bounded intensity oscillations appear if the $\mathcal{PT}$ symmetry of the Hamiltonian is unbroken, where
all quasienergies are real (see Fig. \ref{fig5}(b)). Nonstationary light propagations
of unbounded intensity oscillations appear if the $\mathcal{PT}$ symmetry of the Hamiltonian is broken, where at least one of
the quasienergies is complex (see Fig. \ref{fig5}(a)).
Therefore, the propagation dynamics of our modulated system (1) also can be manipulated by tuning the modulation parameter.

\section{Optical realization}
Below, we discuss the experimental possibility of observing
our theoretical predictions. Recently, several
$\mathcal{PT}$-symmetric optical systems were experimentally realized
\cite{makris07,makris08,muss08,ruter,guo}. The complex refractive index of gain or
loss effects can be obtained from quantum-well lasers or
photorefractive structures through two-wave mixing \cite{yariv}.
Further, periodic modulations can be introduced by out-of-phase
harmonic modulations of the real refractive index
\cite{longh09,longh12,szameit} or periodic curvature along the propagation
direction \cite{longh09,longh12,della,zeuner}.
In our system, the propagation of a light wave along the $z$ axis obeys  the wave equation for its dimensionless electric field amplitude $\phi(x,z)$:
\begin{eqnarray} \label{waveequation}
i\frac{\partial \phi(x,z)}{\partial z}&=&-\frac{1}{2k}\frac{\partial^2 \phi(x,z)}{\partial x^2}+V(x,z)\phi(x,z)
\end{eqnarray}
with refractive index $V(x,z)=V_R(x,z)+iV_I(x)$. The real part is $V_R(x,z)=V_0(x)+V_1(x,z)$ with the symmetric double-well function $V_0(x)$ and periodic modulation $V_1(x,z)$, which periodically changes the distance between two waveguides. The imaginary part $V_I(-x)=-V_I(x)$ consists of antisymmetric functions, and in order to guarantee the $\mathcal{PT}$-symmetry of our system, we defined $f(z)$ to be a periodic even function.
In our numerical simulations, we chose
\begin{eqnarray} \label{potential}
V_0(x)&=&-\rho\Big(\exp\Big[-(\frac{x+\frac{\omega_s}{2}}{\omega_x})^6\Big]+
\exp\Big[-(\frac{x-\frac{\omega_s}{2}}{\omega_x})^6\Big]\Big), \nonumber \\
V_1(x,z)&=&-\rho\alpha\Big(\exp\Big[-(\frac{x+\frac{f(z)}{2}}{\omega_x})^6\Big]+
\exp\Big[-(\frac{x-\frac{f(z)}{2}}{\omega_x})^6\Big]\Big),
\nonumber \\
V_I(x)&=&-\rho\beta\Big(\exp\Big[-(\frac{x+\frac{\omega_s}{2}}{\omega_x})^6\Big]-
\exp\Big[-(\frac{x-\frac{\omega_s}{2}}{\omega_x})^6\Big]\Big),
\nonumber \\
f(z)&=&\omega_s-\mu A \cos(\omega z).
\end{eqnarray}
Here, ($\rho, \alpha, \beta, \mu$) are four parameters describing the refractive index, and $\omega_x$ is the waveguide width. Further, $f(z)$ denotes the modulation of the refractive index where $\omega_s$ is the distance between two waveguides. Parameters $A$ and $\omega$ are the modulation amplitude and frequency, respectively. Parity operator $\hat{P}: \hat{x}\rightarrow -\hat{x}$ and $\hat{p}\rightarrow -\hat{p}$ has
the effect of reversing the transverse direction.
Time-reversal operator $\hat{T}: \hat{x}\rightarrow \hat{x}, \hat{p}\rightarrow -\hat{p}, i\rightarrow -i$ and $z\rightarrow -z$,
has the effect of reversing the propagation direction. Therefore, the
waveguide would be $\mathcal{PT}$-symmetric if $V(x,z)=V^{*}(-x,-z)$, where the asterisk ``$*$'' represents complex conjugation.

 \begin{figure}[htp]
\center
\includegraphics[width=3.5in]{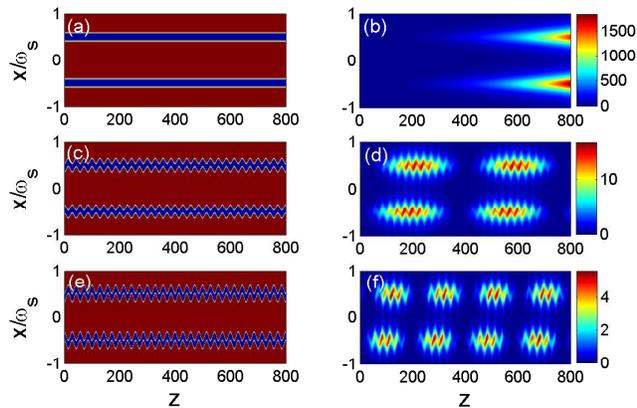}
\caption{(Color online) Light propagation in a $\mathcal{PT}$-symmetric
two-channel coupler. First row: (a) the refractive index distribution
V(x,z) and (b) the light propagation $|\phi(x,z)|^2$ for unmodulated waveguides. Second row: (c) the refractive index distribution
V(x,z) and (d) the light propagation $|\phi(x,z)|^2$ for modulated waveguides with $A/\omega=1.5$. Third row: (e) the refractive index distribution
V(x,z) and (f) the light propagation $|\phi(x,z)|^2$ for modulated waveguides with $A/\omega=2.4$.} \label{fig6}
\end{figure}

To demonstrate how to control $\mathcal{PT}$ symmetry by periodically modulating the distance between two waveguides, we simulated the two-channel coupler by directly integrating the continuous wave equation (\ref{waveequation}). In our numerical simulation,
the initial states were chosen to be the lowest Wannier modes for
isolated individual waveguides, and parameter $k$ was set to
$k=1$. Fixing $\omega_s=3.2$, $\omega_x=0.3$, $\mu=0.4$, $\rho=2.78$, $\alpha=1$, $\beta=0.0252$, and $\omega=0.22$,
we simulated continuous wave equation (\ref{waveequation}) for different modulation amplitudes $A$. As in Ref. \cite{szameit}, $\omega_x$
and $\omega_s$ are in units of 10 $\mu m$, and $\rho=2.78$ corresponds to a real refractive index of $3.1\times10^{-4}$.

According to the analytical analysis in Section III, if the $\mathcal{PT}$ symmetry of the Hamiltonian is unbroken, the light
propagations are stationary oscillations, and nonstationary light propagations of the unbounded intensity oscillations appear if the $\mathcal{PT}$ symmetry of the Hamiltonian is broken. As an analogy,
we numerically explore the light propagation
in a $\mathcal{PT}$-symmetric coupler
under different modulation
amplitudes. In Fig. \ref{fig6}, the left column shows
the refractive index distributions $V(x,z)$
and the right column shows the light propagation $|\phi(x,z)|^2$
obtained by numerically integrating continuous wave equation
(\ref{waveequation}). First, for comparison, in Figs. \ref{fig6}(a) and \ref{fig6}(b) we use the unmodulated system in the region of broken $\mathcal{PT}$ symmetry as a reference system.
Light propagation $|\phi(x,z)|^2$ shows nonstationary oscillations with a growing total intensity for this unmodulated waveguide (see Fig. \ref{fig6}(b)). In contrast to the unmodulated case $A/\omega=0$, the stationary periodic oscillations of light propagation $|\phi(x,z)|^2$ confirm
the existence of $\mathcal{PT}$ symmetry in the modulated cases
$A/\omega=1.5$ and $A/\omega=2.4$ (Figs. \ref{fig6}(d) and \ref{fig6}(f)).
Specifically, when the modulated parameter approaches the zero point of the Bessel function, light propagation $|\phi(x,z)|^2$ demonstrates
more periodic oscillations with propagation behavior that is similar to that of a modulated Hermitian coupler. This means that the unbroken $\mathcal{PT}$-symmetric range can be modulated to span a wider parameter region by periodically modulating the coupling strength of the $\mathcal{PT}$-symmetric optical coupler.
Therefore, our simulated results are
qualitatively consistent with those predicted by coupled-mode equation (\ref{twomodeequation}).
\\

\section{Summary}
In summary, we have investigated how to control the balanced gain and loss in a $\mathcal{PT}$-symmetric optical coupler by periodically modulating the coupling strength between two waveguides. Using the
high-frequency Floquet method, the modulated system is effectively described by an effective averaged system whose gain and loss can be
modulated by adjusting the modulation amplitude or frequency, and such an original non-Hermitian system can even be modulated into an effective Hermitian system.
The spontaneous $\mathcal{PT}$-symmetry-breaking transition was analytically derived and is quite consistent with the numerical simulation. It is revealed that the unbroken $\mathcal{PT}$-symmetric range can be modulated to span a wider parameter region. Furthermore, we discuss the experimental possibility of observing these theoretical results. The simulated results obtained by directly integrating the continuous wave equation are
qualitatively consistent with the ones predicted by the coupled-mode equation. Our results may provide a promising approach for controlling the gain and loss of a realistic system.

\begin{acknowledgments}
We acknowledge helpful discussions with Chaohong Lee.
This work is supported by the NNSFC under Grants No. 11465008, No. 11574405, and No. 11426223, the Hunan Provincial Natural Science Foundation under Grants No. 2015JJ2114, No. 2015JJ4020 and No. 14JJ3114, the Scientific Research Fund of Hunan Provincial Education Department under Grant No. 14A118.

\end{acknowledgments}

\end{CJK*}
\end{document}